\documentclass{article}
\usepackage{amsmath, amsthm, amssymb,mathtools}
\usepackage{times}
\usepackage{enumitem}
\usepackage[boxruled,vlined,linesnumbered]{algorithm2e}

\usepackage[pdftex]{graphicx,color} 

\let\oldnl\nl
\newcommand{\nonl}{\renewcommand{\nl}{\let\nl\oldnl}}
\setlist[enumerate]{itemsep=0mm}
\setlist[itemize]{itemsep=0mm}
\SetAlFnt{\small} 

\newcommand{\hk}{\hspace*{0.5cm}}
\newcommand{\bmp}{\begin{minipage}}
\newcommand{\emp}{\end{minipage}}
\newcommand{\mlog}{\rm {L}og}
\newcommand{\rt}{\textcolor{red}}
\newcommand{\Id}{\mbox{\it Id}}
\newcommand{\ins}{\mbox{\footnotesize\sf insert}}
\newcommand{\last}{\mbox{\footnotesize\sf last}}
\newcommand{\first}{\mbox{\footnotesize\sf first}}
\newcommand{\suc}{\mbox{\footnotesize\sf succ}}
\newcommand{\pre}{\mbox{\footnotesize\sf prec}}
\newcommand{\precede}{\mbox{\footnotesize\sf dprecede}}
\newcommand{\next}{\mbox{\footnotesize\sf dnext}}
\newcommand{\delete}{\mbox{\footnotesize\sf delete}}
\newcommand{\reverse}{\mbox{\footnotesize\sf reverse}}
\newcommand{\weight}{\mbox{\footnotesize\sf weight}}
\newcommand{\modify}{\mbox{\footnotesize\sf add}}
\newcommand{\modifyweight}{\mbox{\footnotesize\sf add-weight}}
\newcommand{\findminweight}{\mbox{\footnotesize\sf find-min-weight}}
\newcommand{\findmaxweight}{\mbox{\footnotesize\sf find-max-weight}}
\newcommand{\searchweight}{\mbox{\footnotesize\sf search-weight}}
\newcommand{\findmin}{\mbox{\footnotesize\sf  find-min}}
\newcommand{\findmax}{\mbox{\footnotesize\sf find-max}}

\newcommand{\findindex}{\mbox{\footnotesize\sf find-rank}}
\newcommand{\findelement}{\mbox{\footnotesize\sf find-element}}
\newcommand{\get}{\mbox{\footnotesize\sf get-value}}
\newcommand{\changesign}{\mbox{\footnotesize\sf change-sign}}
\newcommand{\changesignweight}{\mbox{\footnotesize\sf change-sign-weight}}
\newcommand{\parent}{\mbox{\footnotesize\sf dparent}}
\newcommand{\roo}{\mbox{\footnotesize\sf  droot}}
\newcommand{\cost}{\mbox{\footnotesize\sf dcost}}
\newcommand{\val}{\mbox{\footnotesize\sf dval}}
\newcommand{\dindex}{\mbox{\footnotesize\sf dindex}}
\newcommand{\mincost}{\mbox{\footnotesize\sf dmincost}}
\newcommand{\minval}{\mbox{\footnotesize\sf dminval}}
\newcommand{\maxcost}{\mbox{\footnotesize\sf dmaxcost}}

\newcommand{\update}{\mbox{\footnotesize\sf dupdate}}
\newcommand{\link}{\mbox{\footnotesize\sf dlink}}
\newcommand{\cut}{\mbox{\footnotesize\sf dcut}}
\newcommand{\evert}{\mbox{\footnotesize\sf  devert}}
\newcommand{\expose}{\mbox{\footnotesize\sf  adexpose}}
\newcommand{\construct}{\mbox{\footnotesize\sf  adconstruct}}
\newcommand{\destroy}{\mbox{\footnotesize\sf  addestroy}}
\newcommand{\rotateleft}{\mbox{\footnotesize\sf  adrotateleft}}
\newcommand{\rotateright}{\mbox{\footnotesize\sf  adrotateright}}
\newcommand{\conceal}{\mbox{\footnotesize\sf  adconceal}}

\newcommand{\tailp}{\mbox{\it tail+}}
\newcommand{\head}{\mbox{\it head}}
\newcommand{\tail}{\mbox{\it tail}}
\newcommand{\Hash}{H}
\newcommand{\searchcost}{\mbox{\footnotesize\sf dsearchcost}}
\newcommand{\minuscost}{\mbox{\footnotesize\sf dminuscost}}
\newcommand{\minusindex}{\mbox{\footnotesize\sf dminusindex}}
\newcommand{\minusval}{\mbox{\footnotesize\sf dminusval}}
\newcommand{\searchindex}{\mbox{\footnotesize\sf dsearchindex}}
\newcommand{\n}{[n]}
\newcommand{\hs}{\hspace*{0.6cm}}

\newcommand{\preftr}{\mbox{\footnotesize\sf preftr}}
\newcommand{\prefrv}{\mbox{\footnotesize\sf prefrv}}
\newcommand{\tr}{\mbox{\footnotesize\sf tr}}
\newcommand{\rv}{\mbox{\footnotesize\sf rv}}
\newcommand{\bint}{\mbox{\footnotesize\sf bi}}
\newcommand{\transrev}{\mbox{\footnotesize\sf transrv}}

\newcommand{\Query}{\mbox{\footnotesize\sf Query}}
\newtheorem{thm}{Theorem}
\newtheorem{cor}{Corollary}
\newtheorem{fait}{Claim}
\newtheorem{rmk}{Remark}
\newtheorem{defin}{Definition}
\newtheorem{ex}{Example}
\newtheorem{pb}{Problem}

\newcommand{\bp}{\begin{pb}\rm}
\newcommand{\ep}{\end{pb}}
\newcommand{\br}{\begin{rmk}\rm}
\newcommand{\er}{\end{rmk}}
\newcommand{\bdefin}{\begin{defin}\rm}
\newcommand{\edefin}{\end{defin} }
\newcommand{\bex}{\begin{ex}\rm}
\newcommand{\eex}{\end{ex}}

\newcommand{\bthm}{\begin{thm}}
\newcommand{\ethm}{\end{thm}}
\newcommand{\bcor}{\begin{cor}}
\newcommand{\ecor}{\end{cor}}
\newcommand{\bfn}{\begin{fait}}
\newcommand{\efn}{\end{fait}}

\renewcommand{\Box}{\rule{1.5mm}{3mm}}

\begin{document}

\begin{center}

{\bf\large $\mlog$-Lists and Their Applications to Sorting by Transpositions,\\ Reversals and Block-Interchanges}

%

\vspace*{1cm}

Irena Rusu\footnote{Irena.Rusu@univ-nantes.fr}

L.I.N.A., UMR 6241, Universit\'e de Nantes, 2 rue de
la Houssini\` ere,\\

 BP 92208, 44322 Nantes, France
\end{center}



\vspace*{1cm}

\hrule
\vspace{0.3cm}

\noindent{\bf Abstract} 

Link-cut trees have been introduced by D.D. Sleator and R.E. Tarjan (Journal of Computer and System Sciences, 1983) 
with the aim of efficiently 
maintaining a forest of vertex-disjoint dynamic rooted trees under {\em cut} and {\em link} 
operations. These operations respectively disconnect a subtree from a tree, and join two trees
by an edge. Additionally, link-cut trees allow to change the root of a tree and to perform 
a number of updates and queries on cost values defined on the arcs of the trees. All these
operations are performed in $O(\log\, n)$ amortized or worst-case time, depending on the
implementation, where $n$ is the total size of the forest.

In this paper, we show that a list of elements implemented using link-cut trees (we call it a {\em $\log$-list}) allows us
to obtain a  common running time of $O(\log\, n)$ for the classical operations on lists, but also
for some other essential operations that usually take linear time on lists. Such operations require to 
find the minimum/maximum element in a sublist defined by its endpoints, the position of a given element in the list or 
the element placed at a given position in the list; or they require to add a value $a$, or to multiply by $-1$, all
the elements in a sublist.

Furthermore, we use $\log$-lists to implement several existing algorithms for sorting permutations by
transpositions and/or reversals and/or block-interchanges, and obtain $O(n\,\log\, n)$ running time for all
of them. In this way, the running time of several algorithms is improved, whereas
in other cases our algorithms perform as well as the best existing implementations.

\medskip

\noindent {\bf Keywords:} efficient data structure; double-linked list; link-cut tree; dynamic list ranking; permutation sorting 
\vspace{0.2cm}

\hrule

\section{Introduction}
Many data structures have been defined up to now, allowing to store, modify  and query a set of
elements (see \cite{cormen2009introduction} for a non-exhaustive introduction). Each of these data structures has specificities related to the type of the set to
be stored (disjoint elements or not, ordered elements or not), but overall the
operations to be performed as efficiently as possible are: insert an element, delete an element, find an element,
find the maximum/minimum element, find the predecessor/successor of an element (if it is defined), modify 
(the identifying key of) an element.

When the set is ordered (then we call it a {\em list}), an important number of applications exists where several consecutive 
elements must be inserted/deleted/modified simultaneously.  The above-mentioned  data structures applied to
the ordered case allow only a sequential treatment of each element. Only linked-lists allow to 
insert or delete consecutive elements in/from the sets they represent, but the absence
of a  sublinear worst-case time access to an element given by its position makes that all the other 
operations on the linked-lists  are too time-consuming.

In this paper, we use link-cut trees introduced in \cite{sleator1983data} to define a data structure,
that we call a $\log$-list, on which a significant number of useful operations take $O(\log\, n)$ time.
These operations include the classical operations on lists (delete, insert a sublist), but also, for
instance, simultaneously adding a real value to the values of a sublist, finding the minimum value in a sublist,
or finding the element at the $i$-th position in the list (see the next section for more precisions).
We subsequently use $\log$-lists to improve the running time of several existing algorithms for sorting a permutation by 
(selected types of) transpositions and/or  reversals and/or block-interchanges.

The paper is organized as follows. In Section \ref{sect:log}, we present the operations to be performed on $\log$-lists 
and the implementation of $\log$-lists as link-cut trees. In Section \ref{sect:linkcuttrees}, we recall the general
features of link-cut trees, and propose additional features. In Section \ref{sect:listoper}, we show that 
$\log$-lists achieve $O(\log\, n)$ running time for all the operations we propose for them, including
when the elements have weights and the operations are performed on the weights rather than on the elements.
In Section \ref{sect:applications} we give the aforementioned applications to permutation sorting.
Section \ref{sect:conclusion} is the conclusion.

\section{$\mlog$-lists}\label{sect:log}

Let $L=(x_1, x_2, \ldots, x_n)$ be an ordered set (or {\em list}) of not necessarily distinct elements with values from a numerical set $\Sigma$.
We wish to perform in $O(\log\, n)$ time (and less when possible) the operations below, termed {\em list-operations}, on the list $L$.
We assume each {\em element} is given by a pointer to it. An element is therefore seen as a cell containing {\em the value} of the element.
Thus having (a pointer to) an element and getting the value of the element are two distinct requests. The list-operations are:

\begin{itemize}
 \item $\first(L)$, which returns (a pointer to) the first element in (non-empty) $L$, or null if $L$ is empty
 \item $\last(L)$, which returns (a pointer to) the last element in (non-empty) $L$, or null if $L$ is empty
 \item $\get$(list $L$, element $x$), which returns the value of the element $x$ of $L$.
 \item $\suc$(list $L$, element $x$), which returns the value of the element immediately following $x$ in $L$ if $x\neq \last(L)$
 \item $\pre$(list $L$, element $x$), which returns the value of the element immediately preceding $x$ in $L$, if $x\neq \first(L)$
 \item $\ins$(list $L$, list $L_1$, element $x$) which inserts a list $L_1$ in (non-empty) $L$ immediately after the element $x$
 (similarly: before $x$), and returns $L$. 
 \item $\delete$(list $L$, element $x$, element $y$) which deletes the sublist $L_1$ of $L$ defined by its first element $x$ and its last
 element $y$, and returns $L$ and $L_1$.
 \item $\reverse$(list $L$, element $x$, element $y$) which reverses the order of the elements in the  sublist of $L$ defined by its first element $x$ and its last
 element $y$, and returns the new list $L$.
 \item $\findmin$(list $L$, element $x$, element $y$) which returns an occurrence of the minimum value
in the sublist of  $L$ defined by its first element $x$ and its last element $y$.
 \item $\findmax$(list $L$, element $x$, element $y$) which is similar to $\findmin$ but requires the maximum element.
 \item $\modify$(list $L$, element $x$, element $y$, real $a$) which adds a real value $a$ to the value of each element in the sublist  of $L$ defined by its first element $x$ and its last
 element $y$, and returns the new list $L$. 
 \item $\changesign$(list $L$, element $x$, element $y$) which multiplies by $-1$ the value of each element in the the sublist  of $L$ 
 defined by its first element $x$ and its last element $y$, and returns $L$. 
 \item $\findindex$(list $L$, element $x$) which returns the  value $i$ such that $x$ is the $i$-th element of
 the list $L$ ({\em i.e.} $x=x_i$).
 \item $\findelement$(list $L$, integer $i$) which returns a pointer to the $i$-th element of the list $L$ ({\em i.e.} to $x_i$).
\end{itemize}

When the elements in $L$ have weights (including keys), similar operations may be performed in $O(\log\, n)$ on the weights. 
See Section \ref{sect:keys}.

In a classical double-linked list, the operations $\first$, $\last$, $\get$, $\suc$, $\pre$, $\ins, \delete$ and $\reverse$ need $O(1)$ time, whereas the 
other operations need $O(n)$ time. Our aim is to balance the running times of these operations, seeking 
$O(\log\, n)$ time or less for each of them. The powerful link-cut trees developed by Sleator and Tarjan in \cite{sleator1983data}
in order to deal with dynamic trees apply here in the particular case where all trees are paths (as we show below),
but need additional features that we provide in the next section. 

\begin{figure}
\centering
\resizebox{10cm}{!}{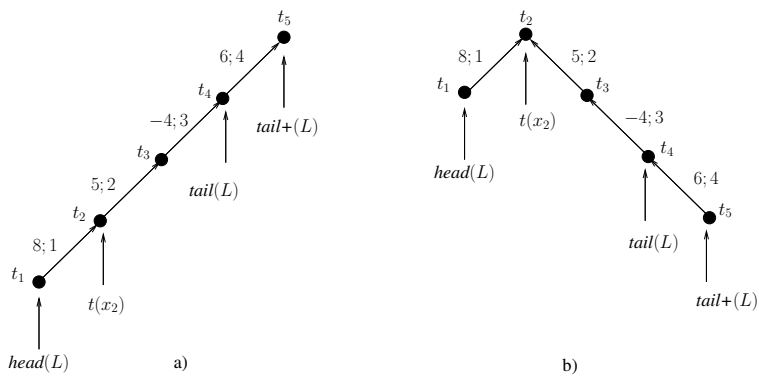}
\caption{\small $\mlog$-list for the list $L=\{8, 5, -4,6\}$. a) link-cut tree $T(L)$ in standard form. b) link-cut tree $T(L)$
in non-standard form. The arc labels are the pairs $val;index$.}
\label{fig:TL}
\end{figure}

We call {\em support} of a directed graph the undirected graph obtained by removing the orientations of the
arcs. Consider the data structure, named a {\em $\log$-list} for $L$ and denoted by $\Hash(L)$, made of:

\begin{itemize}
\item  a rooted tree $T(L)$ whose support is a path, built as follows (see Figure \ref{fig:TL}). It vertex set is  $\{t_1, t_2, \ldots, t_{n+1}\}$ 
and its edges are given by the pairs $(t_i,t_{i+1})$, $1\leq i\leq n$. In its {\sl standard form}, the tree $T(L)$ is 
assumed to have the root $t_{n+1}$, with arcs going towards the root. For each $i$, the arc $e_i$ with endpoints $t_i$ and $t_{i+1}$ has two costs, namely $val(e_i):=x_i$ 
and $index(e_i):=i$.  If the list is empty, then the tree is empty.
\item three pointers $\head(L), \tail(L)$ and $\tailp(L)$ which point respectively to the vertices $t_1, t_n$ and $t_{n+1}$ 
of $T(L)$.  These pointers change only when the list $L$ changes, but are not modified when the root of the tree $T(L)$ changes.

\end{itemize}

\medskip
Obviously, given a list $L$ we can build in $O(n)$ time the tree $T(L)$ and the three pointers $\head(L)$, $\tail(L)$, $\tailp(L)$.
The same data structure is used for all the lists and sublists handled during the aforementioned operations.
Therefore, at each operation, one or several trees must be handled, and they undergo arc deletions and/or arc additions,
that cut or link the existing trees. This forest of trees  is thus implemented using 
{\em link-cut trees} \cite{sleator1983data}, and therefore each tree of it is assumed to be a {\em link-cut} tree
(see the next section for more details). 

\br
Notice that, in $T(L)$, the elements of the list are stored as costs on the {\em arcs}. The vertices in $T(L)$ are
not identified with the elements in $L$. Therefore, when a pointer to an element $x_i$ of $L$ is given,
we assume this means we are provided with a pointer to the source $t(x_i)$ of the arc with cost $x_i$ in the 
standard form of $T(L)$. See Figure \ref{fig:TL}.
\label{rem:pointerbefore}
\er

\br 
Also note that other aggregate operations may be performed similarly to $\modify$, 
since link-cut tree support them, as observed in \cite{tarjan2009dynamic}. See Section \ref{sect:linkcuttrees}
for details about the data structure used by link-cut trees.
\er

\section{Link-cut trees}\label{sect:linkcuttrees}

\subsection{General features}\label{sect:cutlink}

A link-cut tree is a  rooted tree, whose arcs are supposed to be directed towards
the root, so that if $(v,w)$ is an arc then $w$ is the parent of $v$. The following {\em dynamic tree operations} 
may be performed on link-cut trees, in any order 
(where $cost$ is a cost with real values, on the arcs of the link-cut trees in the forest). Each operation takes
$O(\log n)$ time \cite{sleator1983data}.

\begin{enumerate}
 \item $\parent$(vertex $v$), which returns the parent of $v$ in the tree containing it, or null if $v$ is the root. 
 \item $\roo$(vertex $v$), which returns the root of the tree containing $v$.
 \item $\cost$(vertex $v$), which returns $cost(v,\parent(v))$, provided $v$ is not the root of the tree containing it.
 \item $\mincost$(vertex $v$), which returns the vertex $w$ closest to $\roo(v)$ such that $cost(w,\parent(w))$ is minimum
 among all vertices $w'$ on the path from $v$ to $\roo(v)$. Again, it is assumed that $v\neq \roo(v)$.
 \item $\update$(vertex $v$, real $a$), which adds $a$ to the cost of all edges on the path from $v$ to $\roo(v)$.
 \item $\link$ (vertex $v$, vertex $w$, real $a$), which assumes that $v=\roo(v)\neq \roo(w)$ and adds an arc $(w,v)$
 with cost $a$, thus combining the trees containing $v$ and $w$.
 \item $\cut$(vertex $v$), which assumes that $v\neq \roo(v)$ and cuts the arc between $v$ and $\parent(v)$,
 thus dividing the tree initially containing $v$ into two trees.
 \item $\evert$(vertex $v$), which reverses the direction of all arcs on the path from $v$ to $\roo(v)$, thus
 making $v$ the root of the tree.

\end{enumerate}
 
\br Given that the operations $\parent$, $\roo$, $\cost$, $\mincost$ and $\update$ are static (they do not change the forest
of link-cut trees), several costs may be simultaneously defined and used in an arbitrary order.  
Naturally, $\cut$ and $\link$ have a number of cost parameters equal to the
number of cost functions defined on the arcs of the tree.
\label{rem:multiple}
\er

\br 
Also note that $\cost(v)$ takes $O(\log\, n)$ time, and not $O(1)$ time as we could expect. This is due
to the fact that the cost values are not directly stored, but computed using additional information, in order
to allow simultaneous modifications using $\update$  (see Section \ref{sect:add}).
\label{rem:cost}
\er

To achieve $O(\log\, n)$ running time for all these operations, in  \cite{sleator1983data} 
the arc set of each link-cut tree in the forest is partitioned into {\em solid} arcs and {\em dashed arcs}.
Each vertex has at most one ingoing solid arc and, since it has at most one parent, at most one
outgoing solid arc. Thus the {\em solid paths}, which are all the maximal paths formed by solid arcs, partition
the vertex set of the link-cut tree (assuming that a vertex belonging to no solid arc defines alone a trivial
solid path). Each solid path is then represented as a binary tree of height $O(\log\, n)$ 
whose internal nodes represent the arcs of the solid path (with their costs) and whose leaves represent the vertices of the path, in 
such a way that a symmetric traversal of the binary  tree results into a ``spelling'' of the solid path from 
its head to its tail,  including both its vertices and its arcs. The binary trees of all  solid paths of a link-cut tree, which 
also contain a lot of additional information not described here, are then connected in order to depict the 
structure of the link-cut tree. Note that, following \cite{sleator1983data}, we use the
term {\em vertex} for the link-cut trees, and the term {\em node} for the binary  trees.

This structure (see Figure \ref{fig:representation} for a summary) has the twofold advantage of 
being highly parameterizable (the type of the binary  tree,
the definition of solid and dashed arcs) and of being able to reduce operations on link-cut trees to operations
on paths, the later ones being themselves reduced to operations on binary  trees. Then, when a
dynamic operation on a tree has to be performed: either it is a basic operation that may be performed
by querying the binary trees representing the tree without modifying them, and thus without modifying the
solid paths (this is the case of $\parent$ and $\cost$); or it is a complex operation requiring first that 
a solid path be built from the vertex $v$ to the root of its tree (this is the case of $\roo$, $\mincost$, $\update$,
$\cut$, $\evert$, with an exception for $\link$ where the path starts in $w$ instead of $v$).
Building the solid path from $v$ to the root of the tree, and thus the binary tree associated with it, is 
done by an operation called $\expose(v)$, where {\footnotesize\sf a} stands for auxiliary. 
Once $\expose$ is performed, finishing the treatment required by  $\roo$, $\mincost$, $\update$ and $\evert$ needs 
only to move inside the binary  tree, querying it or modifying values. However, the two remaining operations 
$\link$ and respectively $\cut$ need to combine the binary  tree obtained by $\expose$ with another one, and respectively to cut 
it into two trees. Overall, the topological modifications of binary trees are due to
$\expose$, $\link$ and $\cut$, and are implemented using the four auxiliary  operations below:

\begin{itemize}
\item $\construct$(node $r$, node $s$, real $x$), which combines two binary  trees with roots $r$ and
$s$ into another binary  tree with root node having cost $x$, left child $r$ and  right child $s$. 
\item $\destroy$(node $r$), which splits the binary  tree with root $r$ into the two subtrees
with roots given by its left and right child, and returns the two subtrees as well as the cost at node $r$ before splitting.
\item $\rotateleft$(node $r$), which assumes that $r$ has a right child $c$ and performs a left rotation on $r$,
{\em i.e.} $r$ becomes the left child of $c$, whose left child becomes the right child of $r$.
The operation returns the new root of the binary  tree.
\item $\rotateright$(node $r$), which is similar to $\rotateleft$(node $r$), with left and right sides exchanged.
\end{itemize}

The nodes of the binary tree store considerable information allowing to perform these four auxiliary 
operations in constant time, independently of the type of the binary  tree. However, in order to
perform the other dynamic tree operations in $O(\log\, n)$ time (including $\expose$), the type of the
binary tree must be carefully chosen. With locally biased binary trees, amortized $O(\log\, n)$ running time
is achieved. With globally biased binary trees, worst-case  $O(\log\, n)$ running time is achieved 
if in addition the solid arcs are specifically defined as being the {\em heavy arcs} of the link-cut tree.
An arc $(v,w)$ of a link-cut tree is  {\it heavy} if $2size(v)>size(w)$, where $size(u)$ denotes the number of vertices in the 
subtree of $u$, including $u$. The operations defined above remain valid, with the only difference that
when an operation modifying the set of solid paths of a link-cut tree is performed ({\em i.e.} $\roo$, $\mincost$, $\update$, $\evert$, 
$\link$ and $\cut$, which use $\expose$),
then it must be followed by a corrective procedure called $\conceal$ that transforms the (possible temporarily non-heavy) 
solid paths into heavy paths. The efficient implementation of $\conceal$ needs to augment again the data structure,
with data whose update does not modify the running times of the other operations. 

\br
In our description, we assume link-cut trees use heavy paths and (locally or globally) biased binary trees, in order
to achieve the $O(\log\, n)$ amortized or worst-case running time. However, we do not have to go into these details
to explain the additional features we add to the standard link-cut trees data structure, so that we
only use the terms solid paths and binary trees to give our description. 
\er

\begin{figure}[t]

\centering
\begin{tabular}{|l|l|l|l|l|}
\hline
{\bf Configuration}&{\bf Representation} &{\bf Operation}&{\bf Running time}&{\bf Source}\\ \hline
Link-cut tree&Set of solid paths&$\parent(v)$ &$O(\log\, n)$&\cite{sleator1983data}\\ 
&&$\roo(v)$ &$O(\log\, n)$&\cite{sleator1983data}\\
&&$\cost(v)$ & $O(\log\, n)$&\cite{sleator1983data}\\
&&$\mincost(v)$ &$O(\log\, n)$ &\cite{sleator1983data}\\
&&$\update(v,a)$ &$O(\log\, n)$ &\cite{sleator1983data}\\
&&$\link(v,w,a)$  &$O(\log\, n)$&\cite{sleator1983data}\\
&&$\cut(v)$  &$O(\log\, n)$&\cite{sleator1983data}\\
&&$\evert(v)$  &$O(\log\, n)$&\cite{sleator1983data}\\
&&$\searchcost(v)$ &$O(\log\, n)$ & Sect. \ref{sect:add}\\
&&$\minuscost(v)$ &$O(\log\, n)$ & Sect. \ref{sect:add}\\ 
&&$\expose(v)$  &$O(\log\, n)$& \cite{sleator1983data}\\ 
&&$\conceal(p)$  &$O(\log\, n)$& \cite{sleator1983data}\\ \hline
Solid path &Binary tree& path operations&&\\ 
&(abstract)& (not needed here)&&\\ \hline
Binary tree& Locally biased binary tree&$\construct(r,s,x)$& $O(1)$ & \cite{sleator1983data}\\
(abstract)&Globally biased binary tree&$\destroy(r)$& $O(1)$ &\cite{sleator1983data}\\ 
&&$\rotateleft(r)$&  $O(1)$&\cite{sleator1983data}\\
&&$\rotateright(r)$& $O(1)$&\cite{sleator1983data}\\ \hline
\end{tabular}
\caption{\small Link-cut trees and their  representation levels. Only the operations we refer to in this paper
are recorded. The $O(\log\ n)$ running time is the amortized running time for locally biased binary trees,
and the worst-case running time for globally biased binary trees.}
\label{fig:representation}
\end{figure}

\subsection{Additional features}\label{sect:add}

In this section, we propose several modifications of the data structure presented above, in order
to allow the following additional operations on a link-cut tree:

\begin{itemize}
\item[9.] $\searchcost$(vertex $v$, real $a$), which searches for a vertex $w$ on the path from $v$ 
to $\roo(v)$ such that $cost(w, \parent(w))=a$ , assuming the costs are strictly increasing as we go up from $v$ to $\roo(v)$. The operation
returns $w$, if it exists, or the vertex $w'$ with the largest value $cost(w', \parent(w'))$ smaller
than $a$, 
if such a $w'$ exists. Otherwise, it returns $\roo(v)$.
\item[10.] $\minuscost$(vertex $v$), which multiplies by $-1$ all the costs on the path from $v$ to $\roo(v)$.
\end{itemize}

\begin{figure}
\centering
\resizebox{\textwidth}{!}{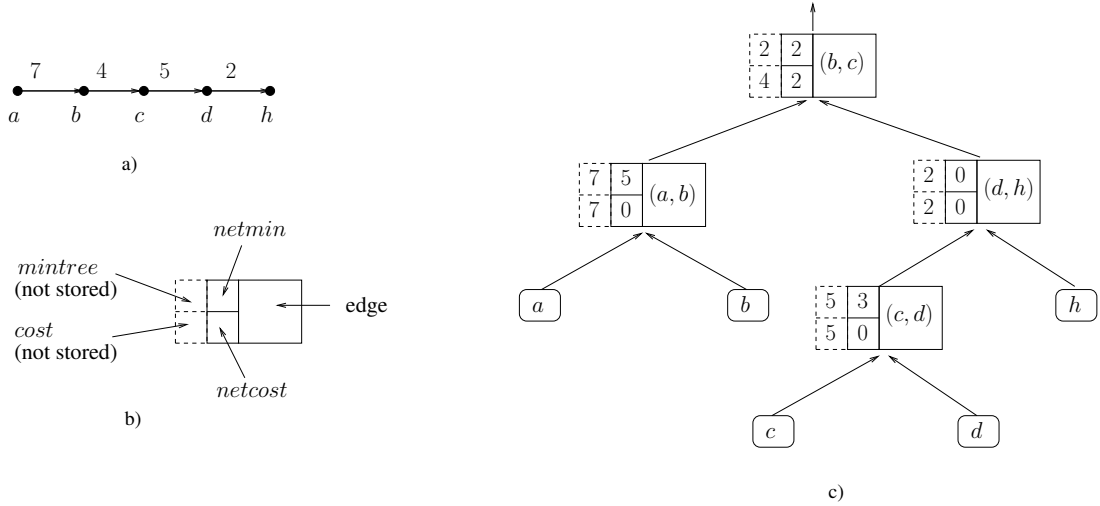}
\caption{\small Binary tree for a given solid path, showing only the essential information related to costs (in particular, only
the $bparent$ pointers are shown, whereas pointers from the parent to its children also exist). a) The solid
path $abcdh$ and the costs on its edges. b) The information stored in each node (cells drawn with plain lines),
and the information not stored, but computed with the help of stored information (cells drawn with dotted lines). c) The
tree built for the solid path in a). }
\label{fig:binary}
\end{figure}

As usual, both these operations start with a call to  $\expose(v)$, which builds the solid path from
$v$ to $\roo(v)$ and the binary tree $B_T$ associated with it. In $B_T$ and as described in
\cite{sleator1983data}, each internal node $e$ (recall it is an arc of $T$) stores, among other information, 
pointers $bparent(e)$ to the parent of $e$ in $B_T$ and $bleft(e),bright(e)$ to respectively the left and right 
child of $e$ in $B_T$. Moreover (see Figure \ref{fig:binary}), it stores two values named $netcost(e)$ and $netmin(e)$, which are related to $cost(e)$ and 
$mintree(e):=\min\{cost(f)\,|\, f\, \mbox{belongs to the subtree rooted at}\, e\}$
by the following equations \cite{sleator1983data}:
\medskip

$netcost(e)=cost(e)-mintree(e)$

\begin{equation*}
netmin(e)=\left\{
  \begin{array}{ll}
   mintree(e)& \mbox{if} \, e\, \mbox{is the root of}\, B_T \hspace*{2.3cm}\\
   mintree(e)-mintree(bparent(e))  & \mbox{otherwise}\\
    
  \end{array}
\right.
\end{equation*}


\medskip

Then $mintree(e)$ is equal to the sum of the $netmin$ values on the path in $B_T$ from $e$ (included) to the root of $B_T$ (included), and
$cost(e)$ is the sum of $mintree(e)$ and $netcost(e)$. Therefore $mintree(e)$ and $cost(e)$ {\em are not}
stored in the tree, but only computed when needed. The values $netcost(e)$ and $netmin(e)$ are initialized 
when the forest of link-cut trees is initialized, in linear time; they are further updated  in $O(1)$ 
when the  link-cut trees and/or their solid paths are modified, by the operations handling these
modifications, namely $\construct, \destroy, \rotateleft$ and $\rotateright$. Note that updating 
the value of $netmin$ at the root of $B_T$ results into an update of $mintree(e)$ and $cost(e)$ in the entire tree, in $O(1)$ time.
\medskip

\noindent{\bf Example.} The value $mintree(c,d)$, which is $5$,  is computed as $netmin(c,d)+netmin(d,h)+netmin(b,c)=3+0+2=5$.
The value $cost(c,d)$, which is also $5$, is computed as $netcost(c,d)+mintree(c,d)=0+5=5$.

\medskip



We are now ready to prove the following claim.

\bfn
Let $T$ be a link-cut tree, and $v$ one of its vertices. Assume the values of the function $cost$ are 
strictly increasing when going from $v$ to $\roo(v)$. Then $\searchcost$(vertex $v$, real $a$) may
be implemented in $O(\log\,n)$ time.
\label{claim:searchcost}
\efn

{\bf Proof.} Perform $\expose(v)$ and let ${v_1, v_2, \ldots, v_p}$, with $v_1=v$, $v_p=\roo(v)$ and $p\geq 2$, be the
vertices on the solid path going from $v$ to $\roo(v)$, in this order. 
Then, by hypothesis, for each arc $(v_i,v_{i+1})$, $1\leq i\leq p-2$, we have 
$cost(v_i,v_{i+1})<cost(v_{i+1},v_{i+2})$. 

Let $B_T$ be the binary tree associated with the solid path, and recall that a symmetric
traversal of $B_T$ allows us to ``spell'' the solid path from its head to its tail, including as well its vertices (which are
the leaves of $B_T$) and its arcs (which are the internal nodes of $B_T$). Then, when the costs of the arcs are
listed by the same symmetric traversal, they are in strictly increasing order, meaning that $B_T$ is a binary search tree.
We deduce that a classical search for $a$ in $B_T$ allows us to find the arc $e$ with cost $a$ (if any), and thus the sought vertex $v_i$ 
(which is the rightmost leaf in the left subtree of $e$). If such an arc $e$ is not found, then again a classical search
allows us to find the arc with largest cost lower than $a$, and the sought vertex $v_j$. 
This search takes a time proportional with the height of $B_T$, that is $O(\log\, n)$.

However, this approach assumes the cost of each arc in $T$ is known. 
Unfortunately, computing $cost(v, \parent(v))$ (which is $\cost(v))$ takes $O(\log\, n)$ time, as indicated in
Remark \ref{rem:cost}, and thus computing all the costs of the arcs would take $O(n\log\, n)$ time. We therefore
need to go deeper into the representation of the costs in the binary tree $B_T$, in order to
reduce the running time to $O(\log\, n)$. Recall that in a binary search tree we only need to compare the
value we are looking for (here, $a$) with the values belonging to a unique branch of the tree, meaning that we only have
to compute the costs of the arcs of $T$ encountered in $B_T$ during this branch traversal. If we show that all these $O(\log\, n)$
costs are computed in $O(\log\, n)$ time, then we are done.

Now, it is easy to see that the $cost$ values of the arcs $e$ of $T$ encountered in $B_T$ during the search 
of $a$ may be computed in $O(1)$ time each, when we  go down this branch. For this, it is sufficient
to notice that $mintree(e)=mintree(bparent(e))+netmin(e)$, except for the root, and that $cost(e)$ is computed
in $O(1)$ time using $mintree$ and $netcost$. Then, all the $cost$ values on the traversed branch of $B_T$ are computed 
in $O(\log\, n)$ time.$\Box$.
\bigskip

We focus now on the second operation we wish to add, $\minuscost$. Again, perform $\expose(v)$ and build 
the binary tree $B_T$ corresponding to the solid path from $v$ to $\roo(v)$. Notice that $\update(v,x)$ updates the 
costs of all the nodes in $B_T$ (and thus of all the arcs on the solid path) in $O(1)$ time by adding $x$ to 
$netmin(r)$, where $r$ is the root node of $B_T$. Several $\update$ operations may be performed consecutively, and each of them has
an immediate effect on $netmin(r)$, implying that we do not have to store the real values involved
in each such operation and, moreover, that we may perform up-down computations by accumulating 
$netmin$ values as in the proof of Claim \ref{claim:searchcost}.  On the contrary, if one wants to 
introduce a multiplicative type of update, one has to store the multiplicative value $y$, since its effect
cannot be reduced to a multiplication of $netmin(r)$ by $y$. If several successive updates
hold, both additive and multiplicative, then all the real values involved in these updates must be stored.
Moreover, the up-down computations become inefficient, since at each level one has to compute all the
stored updates.

Therefore, $\minuscost$ is limited to multiplications by $-1$. In this case, we are able to
ensure an immediate effect on the root of the tree.

\bfn
The link-cut tree data structure may be modified such that, additionally to the other dynamic tree operations,
$\minuscost$(vertex $v$) takes $O(\log\, n)$ time, for each vertex $v$. The space requirements are still in $O(n)$.
\efn

{\bf Proof.} As $\update$ does, $\minuscost$ calls $\expose$ in order to compute the
path from $v$ to $\roo(v)$ and its associated binary tree. We modify the structure of
the binary trees in order to enable efficient sign changes.

Consider the initial state of the forest of link-cut trees, in which the solid
paths of each tree have been defined, and the binary trees to store them are about to be initialized. 
The idea of the proof is to store redundant information in each node $e$ of each binary tree, so that 
each of $cost(e)$ and $-cost(e)$ may be computed using its own series of
$netmin$ values. The series computing $cost(e)$ is the {\em positive series}, whereas the
series computing $-cost(e)$ is the {\em negative series}. They are disjoint, and the type
(positive or negative) of each series is stored in the root $r$ of the tree. A multiplication 
by $-1$ of all the costs in $B_T$ then only requires to exchange the positive and negative 
series. 

Formally, we define each node $e$ in the binary tree to have three parts: one of them, denoted $e^0$, contains 
the usual information stored in the node according to \cite{sleator1983data}, except $netcost$ and $netmin$;
another one, denoted $e^1$, contains two real variables $netcost(e^1)$ and
$netmin(e^1)$ and a pointer $up(e^1)$; the third one, denoted $e^2$, contains two real variables $netcost(e^2)$ 
and $netmin(e^2)$, and a pointer $up(e^2)$. It is assumed that $e^0$, $e^1$ and $e^2$  may
be pointed to separately. Pointers $up(e^1)$ and $up(e^2)$ point to 
$bparent(e)^1$ and to $bparent(e)^2$ respectively, or vice-versa, if $e$ is not the root. If $e$
is the root, then one of them points on its own source node (forming a loop) and the other one is null,
according to rules that will be presented below.

\begin{figure}
\centering
\resizebox{\textwidth}{!}{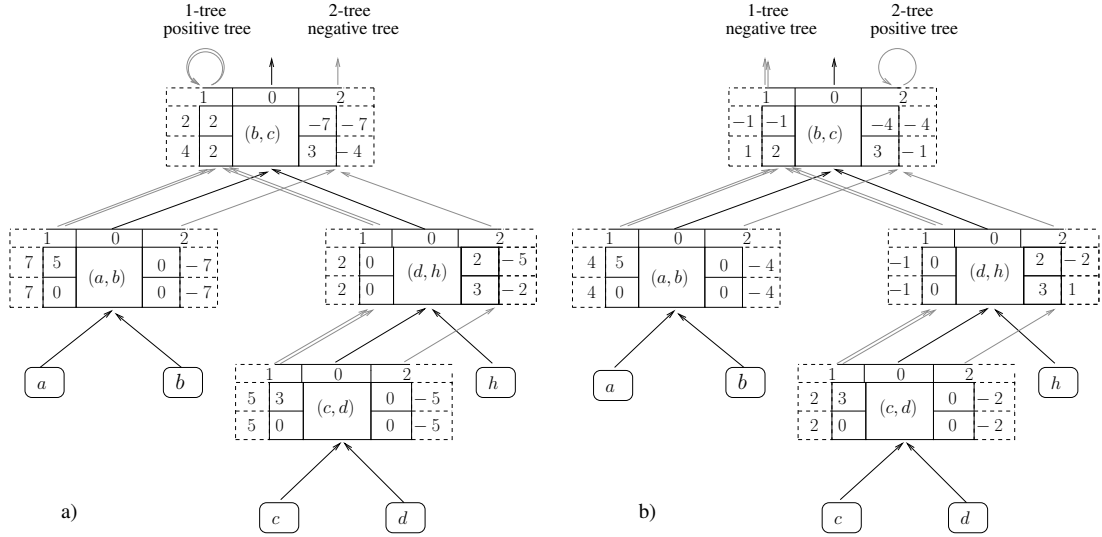}
\caption{\small a) The binary tree for the solid path $abcdh$ in Figure \ref{fig:binary}, with
parts 0, 1 and 2 that are pointed to independently using $up$ pointers (black, double gray and gray arrows respectively). 
b) The same tree after the multiplication
of all values by $-1$ (the up-pointers of the root have changed) and, subsequently, the addition
of the value $3$ to all values in the tree ($+3$ has been added to the $netmin$ value at the
root $r^2$ of the positive tree, and has been subtracted from the $netmin$ value at the
root $r^1$ of the negative tree). The resulting new values of $cost$ and $mintree$ are indicated for
information, but they are not stored (and therefore not computed at this level).}
\label{fig:three}
\end{figure}

Then the (initial) binary tree, whose root is denoted $r$,  may be seen as composed of three 
binary trees (see Figure \ref{fig:three}a): the {\em basic} one given by the 0-parts of the nodes, and the arcs $(e, bparent(e))$;
the {\em 1-tree} given by $r^1$, and the arcs $(e^i, up(e^i))$ such that $up(e^i)=r^1$
or there is a path from $up(e^i)$ to $r^1$; and the {\em 2-tree} given by $r^2$, and the arcs $(e^i,up(e^i))$ such 
that $up(e^i)=r^2$ or there is a path from $up(e^i)$ to $r^2$. These three trees are vertex- and 
arc-disjoint.  The 1-tree and the 2-tree are, in some way, dual to each other, since
one of them computes and updates $cost(e)$, for all $e$, whereas the other one computes and updates
$-cost(e)$, for all $e$. It is understood that the one that computes $cost(e)$, that we call
the {\em positive tree}, is the one whose $up$ pointer forms a loop (see above). The other
one is then the {\em negative tree}. Its $up$ pointer is null.

\medskip 

\noindent{\bf Example.} In Figure \ref{fig:three}a, the three trees are identified by their
root (part $r^0$, $r^1$ or $r^2$ of the global root $r$) and by the arcs forming paths joining this
root. Basically, the 1-tree uses the nodes and arcs on the left (double gray), the 0-tree the nodes and the arcs in the middle (black), 
and the 2-tree the nodes and the arcs on the right (simple gray) of the global tree (This left-middle-right
partition changes when multiplications by $-1$ and topological changes occur, see below.). The 1-tree
is the same as in Figure \ref{fig:binary}. The 2-tree has different values, but it allows us to compute, in the same
way as the 1-tree, the $-cost(e)$ value for each edge $e$ in the solid path. For instance, recall that
we computed $cost(c,d)$ as $netcost(c,d)+mintree(c,d)=0+(3+0+2)=5$. Following the similar
path from the node $(c,d)^2$ ({\em i.e.} the part 2 of the edge $(c,d)$) up to the root $r^2$ ($=(b,c)^2$)
we compute the value $netcost((c,d)^2)+mintree((c,d)^2)=0+(0+2+(-7))=-5$, which is exactly $-cost(c,d)$.

\medskip

The binary trees are initialized simultaneously for the entire forest in its initial state (whatever
this state), as indicated below. Then, they are modified through the operations $\update$, $\minuscost$, as well as 
by the operations $\construct$, $\destroy$, $\rotateleft$ and $\rotateright$ we 
mentioned in Section \ref{sect:cutlink}, which must stay within $O(1)$ running
time. Denote:
\medskip

\hs\bmp[t]{15cm}

$mintree(e^i):=\min\{cost(f^j)\,|\, f^j\,  \mbox{belongs to the subtree rooted at}\, e^i\}$, $i=1,2$ 
\emp\medskip

Then, throughout the modifications above, each binary tree is characterized by the following features:
\medskip

\noindent\bmp[t]{\textwidth}
\begin{itemize}
 \item[A.] Among the 1-tree and the 2-tree, the one with non-null $up$ pointer at its root is the positive tree, {\em i.e.} it computes $cost(e)$ for all
 nodes $e$ in the binary tree; the other one, the negative tree, computes $-cost(e)$ for all nodes $e$ in the binary tree.

 \item[B.] The following equations hold ($i=1,2$):
 \medskip

\hs\bmp[t]{15cm}

 $cost(e^1)=-cost(e^2)$; $|cost(e^1)|=|cost(e^2)|=|cost(e)|$

 $netcost(e^i)=cost(e^i)-mintree(e^i)$

\begin{equation*}
netmin(e^i)=\left\{
  \begin{array}{ll}
   mintree(e^i)& \mbox{if} \, e\, \mbox{is the root of the binary tree} \hspace*{2.6cm}\\
   mintree(e^i)-mintree(up(e^i))  & \mbox{otherwise}\\
    
  \end{array}
\right.
\end{equation*}

\emp\medskip

 \item[C.] Therefore we have:
  \medskip

\hs\bmp[t]{15cm}
 $mintree(e^i)=
 \Sigma_{f^j\, \mbox{\footnotesize belongs to the path from}\, e^i\, \mbox{\footnotesize to the root}}netmin(f^j)$ 
 
 $cost(e^i)=netcost(e^i)+mintree(e^i)$.
\emp\medskip

 \end{itemize}
\emp

In other words, each of the positive and negative trees has the same properties with respect
to the costs as the binary tree used in \cite{sleator1983data}. The need to have two such
trees come from the need to handle both the cost and its negation, which are made
possible by the double choice for the $up$ values.

In the following, we present the initialization step and the aforementioned operations.
Note that A, B and C below are satisfied by the initialization step.
\bigskip

{\it Initialization step.} See Figure \ref{fig:three}a. Let $B$ be a binary tree with root $r$, corresponding to a solid path
of a link-cut tree $T$. The binary tree $B$ is built such that $up(e^i)=bparent(e)^i$ for $i=1,2$,
and $up(r^1)=r^1$, $up(r^2)=null$.

For each node $e$ of $B$, define:
\medskip

\hs\bmp[t]{15cm}
$cost(e^1)=cost(e)$; $cost(e^2)=-cost(e)$

$mintree(e^i):=\min\{cost(f^i)\,|\, f^i\,  \mbox{belongs to the subtree rooted at }\, e^i\}$, $i=1,2.$ 
\emp
\medskip

As in the initial structure in \cite{sleator1983data}, none of these values is stored in $B$.
Only the values $netcost(e^i)$ and $netmin(e^i)$ are stored, and allow to compute them. These
values are defined similarly to \cite{sleator1983data}:
\medskip

\hs\bmp[t]{15cm}
$netcost(e^i)=cost(e^i)-mintree(e^i)$, $i=1,2$

$netmin(e^i)=mintree(e^i)$ if $e$ is the root of the binary tree, and

\hspace*{2cm} $mintree(e^i)-mintree(bparent(e)^i)$, otherwise.
\emp
\medskip


\medskip

\noindent {\bf Example.} In Figure \ref{fig:three}a, the 1-tree (double gray arrows) contains exactly the 
same information as the unique tree in Figure \ref{fig:binary}, by the definitions above. It therefore computes the $cost$ values. The
2-tree (simple gray arrows) similarly computes the $-cost$ values, as it uses the same definitions, but for the values
$-cost(e)$ instead of $cost(e)$.
 For instance, let us see how the values $netcost$ and $netmin$ of the node $(d,h)^2$ are computed. The minimum value in the subtree  with root 
 $(d,h)^2$ of the 2-tree is $-5$ (given by $cost((c,d)^2)$, which is $-cost(c,d)$) and thus $mintree((d,h)^2)=-5$. 
 Similarly, the minimum value in the subtree  with root  $(b,c)^2$ of the 2-tree is $-7$ (given by 
 $cost((a,b)^2)$, which is $-cost(a,b)$) and thus $mintree((b,c)^2)=-7$.  As  $cost((d,h)^2)=-2$ we obtain that 
 $netcost((d,h)^2)=cost((d,h)^2)-mintree((d,h)^2)=-2-(-5)=3$  and $netmin((d,h)^2)=mintree((d,h)^2)-mintree((b,c)^2)=
 -5-(-7)=2$.

\medskip

Then we have:
\medskip

\hs\bmp[t]{15cm}
$mintree(e^i)= \Sigma_{f^i\, \mbox{\footnotesize belongs to the path from}\, e^i\, \mbox{\footnotesize to the root of}}netmin(f^i)$ 

$cost(e^i)=netcost(e^i)+mintree(e^i).$
\emp
\medskip

Consequently, the 1-tree computes the values $cost(e)$ for all nodes $e$ of $B$, whereas the 2-tree
computes the costs $-cost(e)$. The pointers $up(r^1)$ and $up(r^2)$ are correctly initialized.
\bigskip


{\em Modifying the binary tree.} See Figure \ref{fig:three}b. The binary tree, as defined above, is topologically modified by several dynamic tree
operations, which call the four auxiliary operations $\construct, \destroy$, $\rotateleft$ and $\rotateright$.
In our version of the binary tree, each of them is implemented separately for the positive and the negative
trees, using the same methods as in \cite{sleator1983data}. Therefore, the  operation $\construct(r,s,x)$
then applies the (simple) $\construct$ operation in \cite{sleator1983data} once
for the two positive trees and the cost value $x$ for $u^1$, and once for the two negative trees and 
the cost value $-x$ for $u^2$. The operations $\destroy(r)$, $\rotateleft(r)$ and $\rotateright(r)$
also apply twice the same (simple) operations in \cite{sleator1983data}. Then, these operations have the same
running time as those in  \cite{sleator1983data}, that is $O(1)$ time, and the $O(\log\, n)$ running
time of the dynamic operations using them follows as in \cite{sleator1983data}.

Additionally to the topological changes, the operations $\update$ and $\minuscost$ modify some
values of the binary tree. In our version of the binary tree, $\update(v,x)$ first performs $\expose(v)$ and, 
in the binary tree $B$ associated with the solid path from $v$ to $\roo(v)$, adds $x$ to $netmin(r^i)$ and 
removes $x$ from $netmin(r^{3-i})$, where $r^i$ is the root of the positive tree of $B$.
Operation $\minuscost(v)$ first performs $\expose(v)$ and, in the binary tree $B$ associated with the
solid path from $v$ to $\roo(v)$, exchanges the positive and the negative trees by appropriately
modifying the $up$ pointers of their roots.

Obviously, our version of the binary tree only duplicates the operations in the original 
version, and all the running times are unchanged.
$\Box$
 \bigskip

\section{List-operations on $\log$-lists}\label{sect:listoper}
Now, the tree $T(L)$ of any $\log$-list $L$ is seen as a link-cut tree, whose support is a path (see Section \ref{sect:log})
and which has two cost operations, $val$ and $index$,  defined on each of its arcs. The underlying structure ({\em i.e.}
the binary trees associated with its solid paths) are built in $O(n)$ when the $\log$-list is initialized.
%
%

\subsection{Main result}

We show that:

\bthm
In a $\log$-list, all the list-operations take $O(\log\, n)$ time, except for  $\first$ and $\last$ that take $O(1)$ time. 
\label{thm:rev}
\ethm

{\bf Proof.} We assume that, before each operation, $T(L)$ is in standard form (otherwise, we apply $\evert(\tailp(L))$).
We only have to show how dynamic tree operations allow to perform list-operations. Note that
$L$ and $L_1$ are both represented as $\log$-lists with cost values
$val$ and $index$, and therefore support the dynamic tree operations. Recall that when a pointer is
given to an element $L$ in the list, this means a pointer is given to the vertex $t(x)$ of $T(L)$
which is the source of the arc recording $x$ in the standard form of $T(L)$ (according to Remark \ref{rem:pointerbefore}).

Operations $\first(L)$ and $\last(L)$ need only to return $\head(L)$ and
$\tail(L)$ respectively, and take constant time. 

Operation $\get(L,x)$ may be realized by a simple call to $\val(t(x))$, which is a call to $\cost$
when the cost function is called $val$.

Operations $\suc(L,x)$ and $\pre(L,x)$ are easy to implement. For $\suc(L,x)$, since $T(L)$
is in standard form, let $y$ be $\parent(t(x))$. Then the value returned by $\val(y)$ is the successor of $x$ in $L$,
where $\val$ is the variant of the function $\cost$ when the cost is given by the function $val$.
For $\pre(L,x)$, perform $\evert(\head(L))$ (which does not change the pointer $t(x)$) and return
the value $\val(z)$, where $z=t(x)$. Note here that  $t(\suc(x))$ and $t(\pre(x))$ are easy to compute
in $O(\log\,n)$ by respectively returning $y$, and $\parent(z)$. These operations are used in Algorithms
\ref{algo:2}, \ref{algo:3} and \ref{algo:4} below.

\begin{figure}
\centering
\resizebox{0.95\textwidth}{!}{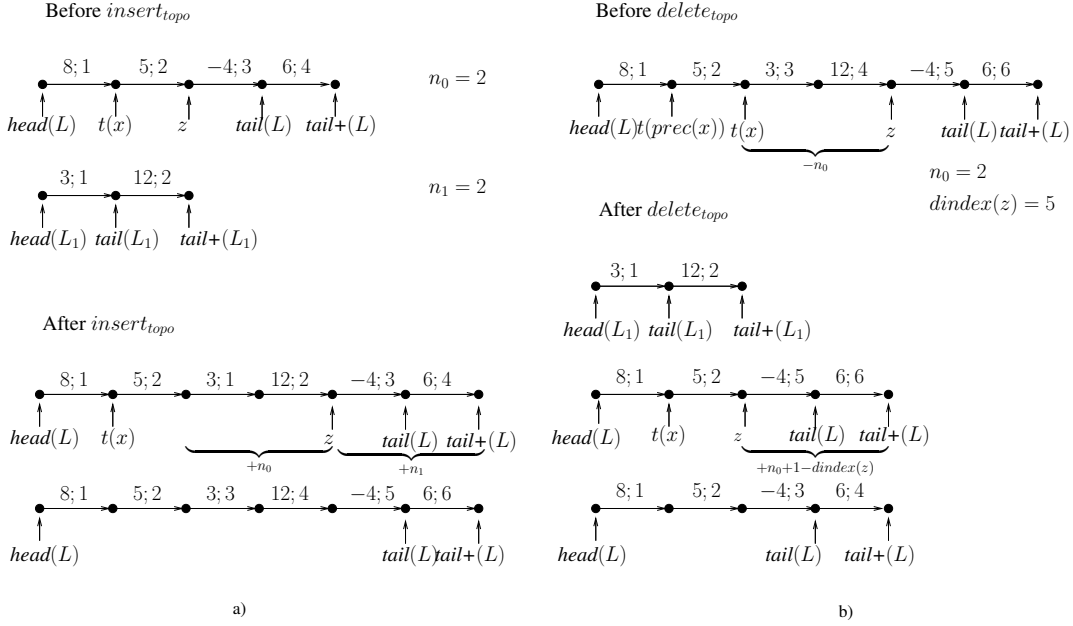}
\caption{\small Index correction procedures for $\ins$ and $\delete$: a) $\ins(L, L_1,x)$ for
$L=\{8, 5, -4, 6\}$, $L_1=\{3, 12\}$ and $x=x_2=5$, using Algorithm \ref{algo:2}; b) $\delete(L,x,y)$
for $L=\{8, 5, 3, 12, -4, 6\}$, $x=x_3=3$ and $y=x_4=12$, using Algorithm \ref{algo:3}. Each arc
indicates its  $val;index$ pair.}
\label{fig:oper}
\end{figure}

Operation $\ins(L, L_1, x)$ (see Figure \ref{fig:oper}a) is written using classical deletions and insertions of arcs, except that 
each arc deletion is implemented using the $\cut$ dynamic tree operation, whereas each arc insertion uses the 
$\link$ dynamic tree operation. Notice that, since the elements in $L_1$ are on the arcs of $T(L_1)$,
$\tailp(L_1)$ is cut from $L_1$ before inserting $L_1$, that is, before appropriately linking it. However,
the values $val$ and $index$ of the former arcs $(t(x), \parent(t(x))$ in $T(L)$ and  $(\tail(L_1), \tailp(L_1))$ in $T(L_1)$ are 
appropriately recorded on the two new linking arcs (the $\link$ operation allows it, assuming it is 
extended so as to have two cost parameters instead of one, according to Remark \ref{rem:multiple}). Once this is
done by the operation $\ins_{topo}(L, L_1,x)$ (not written here), the $val$ values are in the right order. 
It remains to update the $index$ values, so as to ensure that they correctly compute the position of 
each element of $L$ in the list $L$. This is done as in Algorithm \ref{algo:2}. Note that a call to $\dindex$ means a 
call to the $\cost$ dynamic tree operation, when $cost$ is replaced by $index$.

\begin{algorithm}[t]
\label{algo:2}
\caption{{\small\sf insert}$(L,L_1,x)$}

$\Hash(L)\leftarrow$ $\log$-list representing $L$\tcp*[r]{\small\rm assumes $\tailp(L)$ is the root of $T(L)$}

$\Hash(L_1)\leftarrow$ $\log$-list representing $L_1$\tcp*[r]{\small\rm assumes $\tailp(L_1)$ is the root of $T(L_1)$}

$n_0\leftarrow \dindex(t(x))$\tcp*[r]{\small\rm number of elements in $L$ before $x$ and including $x$}

$n_1\leftarrow \dindex(\tail(L_1))$\tcp*[r]{\small\rm number of elements in $L_1$}

$z\leftarrow t(\suc(x))$\; 

$\Hash(L)\leftarrow \ins_{topo}(L, L_1,x)$\tcp*[r]{\small\rm assumes pointers to $t(x)$ and $z$ did not change}

$\evert(z)$; $\update(\parent(t(x)),n_0)$\tcp*[r]{\small\rm updates the indices of the elements coming from $L_1$}

$\evert(\tailp(L))$; $\update(z,n_1)$\tcp*[r]{\small\rm updates the indices of the elements between $z$ and $\tailp(L)$}

return $\Hash(L)$\;
\end{algorithm}

Operation $\delete(L,x,y)$ (see Figure \ref{fig:oper}b) is written similarly, and outputs the list $L$ as well as the list $L_1$ of deleted
elements. Notice that a tail $\tailp(L_1)$ is added to the subtree deleted from $T(L)$ in order to form $T(L_1)$,
allowing its ingoing arc to receive the $val$ cost equal to $y$ and the corresponding $index$ value (as they were
on the arc $(y, \parent(y))$ of $L$). Adding $\tailp(L_1)$ and recording the costs as indicated is done 
using a $\link$ operation with a trivial tree containing only $\tailp(L_1)$ (we assume a sufficient number
of such trees, at most $n$, is present in the forest, in order to allow  such completion operations on
$\log$-lists). To ensure its concision, the $index$ correcting Algorithm \ref{algo:3} is written in the most 
general case, where no list is empty. We assume that the (omitted) $\delete_{topo}$ operation performs the 
topological changes as described. As both lists $L$ and $L_1$ must be output by the $\delete$ algorithm 
(since they are parts of the forest), both of them are updated. 
\smallskip

\begin{algorithm}[t]
\label{algo:3}
\caption{{\small\sf delete}$(L,x,y)$}

\nonl //Note: we also update the values for the deleted sublist $L_1$  

$\Hash(L)\leftarrow$ $\log$-list representing $L$\tcp*[r]{\small\rm assumes $\tailp(L)$ is the root of $T(L)$}

$n_0\leftarrow \dindex(t(\pre(x)))$\tcp*[r]{\small\rm number of elements in $L$, before $x$}

$z\leftarrow t(\suc(y))$\;

$\evert(z)$; $\update(t(x),-n_0)$\tcp*[r]{\small\rm updates the index for the elements that will go into list $L_1$}

$(\Hash(L),\Hash(L_1))\leftarrow \delete_{topo}(L,x,y)$\tcp*[r]{\small\rm assumes pointer to $z$ is still available}

$\evert(\tailp(L))$; 

$\update(z,n_0+1-\dindex(z))$\tcp*[r]{\small\rm updates the indices of the elements between $z$ and $\tailp(L)$}

return $\Hash(L), \Hash(L_1)$\;
\end{algorithm}

\begin{figure}
\centering
\resizebox{0.5\textwidth}{!}{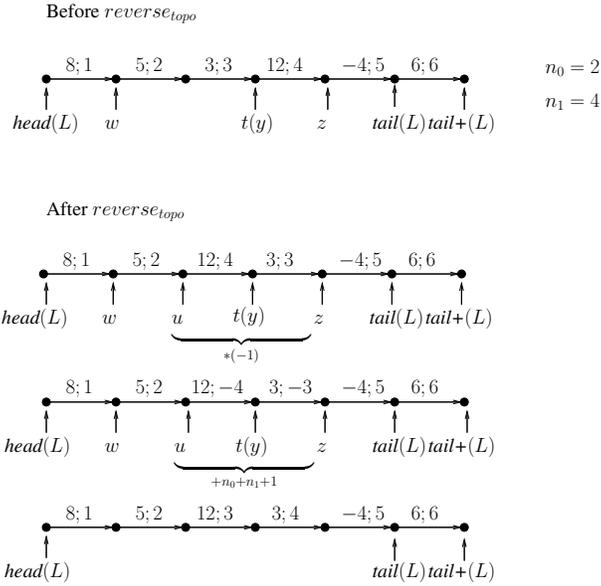}
\caption{\small Index correction procedure for $\reverse(L, x,y)$ with
$L=\{8, 5, 3, 12, -4, 6\}$, $x=x_3=3$ and $y=x_4=12$, using Algorithm \ref{algo:4}. Each arc
indicates its  $val;index$ pair.}
\label{fig:operRev}
\end{figure}

Operation $\reverse(L,x,y)$ (see Figure \ref{fig:operRev}) consists basically in the deletion from $L$ of the sublist with endpoints given
by $x$ and $y$ (which is represented as a $\log$-list and is thus a new tree in the forest), 
its reversal using $\evert(\head(L_1))$ (where $\head(L_1)$ is $t(x)$), the update of the pointers
$\head, \tail$ and $\tailp$, and the insertion of the resulting list in $L$. 
As before, we assume that these changes are realized by the omitted operation $\reverse_{topo}$, and we
only show how to update  the values of $index$. This is done in Algorithm \ref{algo:4} where, for concision reasons,
only the most general case is presented. The operation $\minusindex$ is the same as the operation $\minuscost$ when
the cost is the function $index$.

\begin{algorithm}[t]
\label{algo:4}
\caption{{\small\sf reverse}$(L,x,y)$}

\nonl //Note: we also update the values for the deleted sublist $L_1$  

$\Hash(L)\leftarrow$ $\log$-list representing $L$\tcp*[r]{\small\rm assumes $\tailp(L)$ is the root of $T(L)$}

$w\leftarrow t(\pre(x))$; $z\leftarrow t(\suc(y))$\;

$n_0\leftarrow \dindex(w)$; $n_1\leftarrow \dindex(t(y))$\tcp*[r]{\small\rm number of elements in $L$, before $x$, and up to and including $y$}

$\Hash(L)\leftarrow \reverse_{topo}(L,x,y)$\tcp*[r]{\small\rm assumes pointers to $w,z$ are still available}

$\evert(z)$; $u\leftarrow \parent(w)$;

$\minusindex(u)$\tcp*[r]{\small\rm let the indices in the sublist with endpoints $y$ and $x$ be negative but in increasing order}

$\update(u,n_0+n_1+1)$\tcp*[r]{\small\rm the indices in the sublist with endpoints $y$ and $x$ get the right values}

$\evert(\tailp(L))$;
return $\Hash(L)$
\end{algorithm}

Operation $\findmin(L, x, y)$ first performs a call to $\evert(t(\suc(y)))$, in order to ensure that the arcs of
the path going from $x$ to $\roo(x)$ (which is now $t(\suc(y))$) record the values $val$ corresponding to the
sublist of $L$ between $x$ and $y$. Then $\minval(t(x))$, which is the dynamic tree operation corresponding to $\mincost$ 
when the cost is given by the function $val$, returns the vertex $v$ closest to the root that has minimum value 
$val(v, \parent(v))$. 

Operation $\findmax(L, x, y)$ applies $\findmin(L, x, y)$ once the signs of the elements have been changed
with $\evert(y)$ followed by $\minusval(x)$ (the variant of $\minuscost$ for the cost function $val$), and
returns the opposite of the result. 

Operations $\modify(L, x, y, a)$ and $\changesign(L, x, y)$ are done by simple calls to $\evert(t(\suc(y)))$
and then to $\update(t(x),a)$ and respectively to $\minusval(t(x))$.


Operation $\findindex(L,x)$ only needs to return $\dindex(t(x))$ when $T(L)$ is in standard form, since the cost 
function $index$ is correctly updated. Again, $\dindex$ is the $\cost$ operation when the cost function is $index$.

Operation $\findelement(L,i)$ performs $\evert(\tailp(L))$ in order to ensure that $T(L)$ is  in standard form,
and applies the variant $\searchindex(\head(L),i)$ of $\searchcost$ when the cost function is  $index$.
It is easy to check that the hypothesis of Claim \ref{claim:searchcost} are verified.

Each list-operation uses a constant-bounded number of dynamic tree operations or computations of $t(\suc(x))$ and $t(\pre(x))$
(taking $O(\log\, n)$ time as observed when $\suc$ and $\pre$ were discussed), so the running time 
of each list-operation is bounded by $O(\log\, n)$. $\Box$

\br Note that the traversal of the entire $\log$-list  needs - if we use
the operations $\suc$ and $\get$ on each element - $O(n\, \log\, n)$ time. However,
traversing directly the binary tree associated with the entire list (in its
standard form, and once  $\expose(\head(L))$ has been performed) allows to
visit and modify all the elements in the list in $O(n)$.
\er

%
%


\subsection{$\mlog$-lists with weights}\label{sect:keys}

Now that the underlying data structure of a $\log$-list is fixed, and that the 
large possibilities offered by link-cut trees are understood, it is easy to
see that we can provide one or several weights for each element in $L$
and, for each weight function, add to the list-operations above the following  {\it weighted list-operations}:

\begin{itemize}
 \item $\weight(L,x)$ which returns the weight of a given element $x$ of $L$
 \item $\findminweight(L,x,y)$  which returns the minimum weight of an element  
in the sublist of  $L$ defined by its first element $x$ and its last element $y$
 \item $\findmaxweight(L,x,y)$  which is similar to  $\findminweight(L,x,y)$  but requires the maximum weight
 \item $\modifyweight(L,x,y,a)$ which adds a real value $a$ to the weight of each element in the sublist  of $L$ defined 
 by its first element $x$ and its last element $y$, and returns the new list $L$.
 \item $\changesignweight(L,x,y)$ which multiplies by $-1$ the weight of each element in the sublist  of $L$ defined 
 by its first element $x$ and its last element $y$, and returns the new list $L$.
 \item $\searchweight(L,b)$ which assumes that $L$ is sorted in increasing order of the weights (which are assumed distinct), 
and returns an element of  weight $b$  (if it exists) or a pointer to the element with largest weight lower than $b$ (variant: 
on the element with smallest weight larger than $b$)
\end{itemize}

We easily have:

\bthm
In a $\log$-list with weights, the weighted list-operations above are performed in $O(\log\, n)$ time.
\ethm

{\bf Proof.} The weight of an element $x$ is another cost on the arc $(x,\parent(x))$ in the initial tree $T(L)$
for the list $L$. The dynamic tree operations $\cost$, $\mincost$ (as well as its variant $\maxcost$), $\update$,
$\minuscost$ 
and  $\searchcost$ applied to this new cost allow to perform the  weighted-operations.
$\Box$

\section{Applications}\label{sect:applications}

We give below several applications of $\log$-lists to sorting a permutation by reversals and transpositions.
Best algorithms for these tasks are usually very long and complex, and therefore showing that a new data structure
improves their running times is a fastidious task.  We therefore chose to show: 1) the improvement realized by $\log$-lists 
on recent (and moderately long) algorithms involving prefix/suffix transpositions and reversals (in Subsection \ref{sect:dias}); 2) the
evidence that $\log$-lists generalize permutation trees, that are able to 
efficiently deal with transpositions but not with reversals (in Subsection \ref{sect:perm}). 


\subsection{Terminology}

The definitions below are limited to the results we refer to in the remaining of the section, and do not constitute an exhaustive
list of genome rearrangements. The reader is referred to \cite{fertin2009combinatorics} for a detailed 
survey of genome rearrangements, and results on permutation sorting.

A genome is represented as a permutation (also termed {\em unsigned permutation}) $P=(p_1\, p_2\, \ldots, p_n)$ on $\n:=\{1, 2, \ldots\, n\}$.
In a more precise representation of the genome, the
elements may carry a $+$ or $-$ sign, in which case the permutation is {\em signed}. The inverse permutation 
of $P$ is denoted $P^{-1}$, whereas $\Id$ denotes the identity permutation. 

The {\em transposition} $\tr(P, i,j,k)$, where $1\leq i<j<k\leq n+1$, is the operation that transforms
$P$ into the following permutation:
\medskip

\hs\bmp[t]{15cm}
$P_0:=(p_1\, p_2\, \ldots\, p_{i-1}\, p_j\, p_{j+1}\, \ldots\, p_{k-1}\, \underline{p_i\, p_{i+1}\, \ldots\, p_{j-1}}\, p_k\, \ldots\, p_n)$.
\emp\medskip

In other words, the block of $P$ with endpoints $p_i$ and $p_{j-1}$ is moved between $p_{k-1}$
and $p_{k}$. The {\em reversal} $\rv(P,i,j)$, where $1\leq i\leq j\leq n$, acts differently on unsigned and
signed permutations. When $P$ is unsigned, the reversal is also {\em unsigned} and it transforms $P$
into the following permutation:
\medskip

\hs\bmp[t]{15cm}
$P_1:=(p_1\, p_2\, \ldots\, p_{i-1}\, \underline{p_j\, p_{j-1}\, \ldots\, p_{i+1}\, p_i}\, p_{j+1}\, \ldots\, p_n)$.
\emp\medskip

Equivalently, the order of the elements in the block of $P$ with endpoints $p_i$ and $p_j$ is reversed.
When $P$ is signed, the reversal is also {\em signed} and  and it transforms $P$
into the following permutation:
\medskip

\hs\bmp[t]{15cm}
$P_1:=(p_1\, p_2\, \ldots\, p_{i-1}\, \underline{-p_j\, -p_{j-1}\, \ldots\, -p_{i+1}\, -p_i}\, p_{j+1}\, \ldots\, p_n)$.
\emp\medskip

Equivalently, the order of the elements in the block of $P$ with endpoints $p_i$ and $p_j$ is reversed, and
the signs are changed.
A transposition $\tr(P, i,j,k)$ or a reversal $\rv(P,i,j)$ is {\em prefix} if $i=1$, in which case
they are denoted $\preftr(P, j,k)$ and $\prefrv(P,j)$ respectively. A {\em suffix} transposition or reversal
is defined in a similar way, but involves a suffix of $P$ instead of a prefix.

The {\em block interchange} $\bint(P,i, j, k, l)$, where $1\leq i<j\leq k<l\leq n+1$, transforms $P$ into the following permutation:

\medskip

\hs\bmp[t]{15cm}
$P_2:=(p_1\, p_2\, \ldots\, p_{i-1}\, \underline{p_k\, p_{j+1}\, \ldots\, p_{l-1}}\,  p_{j}\, \ldots\, p_{k-1}\, \underline{p_i\, p_{i+1}\, \ldots\, p_{j-1}}\, p_{l}\, \ldots\, p_n)$.
\emp\medskip

Equivalently, the underlined blocks are switched.

\subsection{Sorting by Prefix/Suffix Transpositions and Reversals in $O(n\,\log\, n)$ time}\label{sect:dias}

Several algorithms in the literature share similar principles for sorting signed or unsigned permutations
either by transpositions only, or by transpositions and reversals, when all operations are assumed to
be prefix or suffix. We present in detail one of them, which allows us to precisely state the bottlenecks
of such an approach in terms of running time. Then we show how to address these bottlenecks with $\log$-lists and 
we conclude on all the similar variants (given in Figure \ref{fig:Table}).

\begin{figure}[t]

\centering
\begin{tabular}{|cc|c|c|c|}
\hline
&&{\small\sc Transpositions} &{\small\sc Unsigned Reversals}&{\small\sc Signed Reversals}\\ 
&&&{\small\sc and Transpositions}&{\small\sc and Transpositions}\\\hline
{\bf Prefix}&Source&\cite{dias2002sorting}& \cite{dias2015sorting}&\cite{lintzmayer2014sortingsigned}\\
&Result& $2$-approx.& asymptotic 2-approx. &asymptotic 2-approx.\\
(Similarly: Suffix)&Best& $O(n \log\,n)$ {\bf (here)} & $O(n \log\,n)$ {\bf (here)} &$O(n \log\,n)$ {\bf (here)}\\ \hline
{\bf Prefix and}&Source&\cite{lintzmayer2014sorting}&\cite{lintzmayer2014sorting}&\cite{lintzmayer2014sortingsigned}\\
{\bf Suffix}&Result& $2$-approx. & asymptotic 2-approx.&asymptotic 2-approx.\\
&Best&$O(n \log\,n)$ {\bf (here)} & $O(n^2)$ \cite{lintzmayer2014sorting} &$O(n \log\,n)$ {\bf (here)}\\ \hline

\end{tabular}
\caption{\small Improvements over existing algorithms, whose $O(n^2)$ running time was optimal before our implementation. 
One cell represents a variant 
of the permutation sorting, allowing all the operations indicated in the column header, in all the versions indicated in the line header. The Source, Result and Best  
lines respectively give the reference of the original algorithm, its approximation ratio, and its best current implementation.
Note that in five cases over six, $\log$-lists easily yield the best implementations, similarly to the implementation we proposed
for Algorithm \ref{algo:1}.}
\label{fig:Table}
\end{figure}

In \cite{dias2015sorting}, the asymptotic 2-approximation
algorithm in Algorithm \ref{algo:1} is presented for sorting an unsigned permutation by prefix transpositions and prefix reversals.
Its running time is of $O(n^2)$. Note that the element $n+1$ is added at the end of the permutation. A {\em strip} of $P$ is 
a sequence $p_i, p_{i+1}, \ldots, p_j$, $j\geq i$, of {\em consecutive integers}
either in increasing order (yielding an {\em increasing strip}) or in decreasing order (yielding a {\em decreasing strip}).
A singleton is, by definition, both an increasing and a decreasing strip. 


\begin{algorithm}
\label{algo:1}
\caption{\small Sorting by Prefix Reversals and Prefix Transpositions \cite{dias2015sorting}}
\KwIn{Permutation $P$, number  $n$ of elements}
\KwOut{Number $d$ of prefix reversals and prefix transpositions needed to sort $P$}

 $d\leftarrow  0$\;
\While {$P \neq \Id$}
    { $i \leftarrow 1$\;
      \While{$|p_{i+1}- p_i| = 1$}
           {$i \leftarrow i + 1$\;}
      \If{$p_1 = 1$}{
           $P \leftarrow \preftr(P,i + 1, n + 1)$\;}
      \Else{
           \nonl // Try to find a prefix transposition extending the strips at both ends of the moved block\;
            $a \leftarrow  P_{p_1-1}^{-1} +1$; $la \leftarrow P_{p_{a}-1}^{-1}+1$; $ra \leftarrow P_{p_{a}+1}^{-1}+1$\;
            $b \leftarrow P_{p_1+1}^{-1}+1$; $lb \leftarrow P_{p_{b}-1}^{-1}+1$; $rb \leftarrow P_{p_{b}+1}^{-1}+1$\;
            \If{$|p_{a-1} - p_{a}| \neq 1$}{ 
               \If{$p_{la} \neq 1$ and $|p_{la-1} - p_{la}| \neq 1$} 
                   {\If{$la<a$} {$P\leftarrow \preftr(P,la, a)$\;}}
               \ElseIf{$p_{ra} \neq 1$ and $|p_{ra-1} - p_{ra}| \neq  1$ } 
                   {\If{$ra<a$}{$P\leftarrow \preftr(P, ra, a)$\;}}
            } \Else{\If{$|p_{b-1} - p_{b}| \neq 1$}
               {\If{$p_{lb} \neq 1$ and $|p_{lb-1} -p_{lb} | \neq 1$ }
                   {\If{$lb<b$}{$P\leftarrow  \preftr (P, lb, b)$\;}}
               \ElseIf{$p_{rb} \neq 1$ and $|p_{rb-1} - p_{rb}| \neq 1$ }
                    {\If{$rb<b$}{$P\leftarrow \preftr (P,rb, b)$\;}}}
               }\Else
                  { \nonl// Try to find a prefix reversal/transposition extending one strip of the moved block\;
                     \If{$p_1 \leq p_i$}{
                           $x\leftarrow P_{p_1-1}^{-1}$\;
                           \If{$p_x = p_{x+1} + 1$} 
                              {$P\leftarrow  \prefrv (P,x - 1)$\;}
                           \Else
                              {$P\leftarrow \preftr (P,i + 1, x + 1)$\;}

                      }\Else{
                            $y \leftarrow P_{p_1+1}^{-1}$\;
                            \If{$p_y = p_{y-1} -1$}
                                  {$P\leftarrow  \preftr (P,i + 1, y + 1)$\;}
                            \Else
                                  {$P\leftarrow \prefrv (P,y - 1)$\;}
                           }
                     }
              }
$d \leftarrow d + 1$\;
}
return $d$\;
\end{algorithm}

{\em Idea of Algorithm \ref{algo:1}.} Given the initial set of strips in the permutation, the
algorithm performs one prefix operation at each step, preferring an operation that performs two
strip concatenations to an operation that performs only one concatenation..
To this end, the position $i$ at the end
of the first strip is identified (steps 4-5). If the first element in $P$ is 1, it is not very
useful immediately, so the whole strip is sent at the end of the permutation (steps 6-7). If the first element $p_1$
is not 1, then the algorithm attempts to move a prefix $(p_1\, p_2\, \ldots\,  p_j)$ with $j\geq i$
between $p_1-1$ and its successor (whose positions are $a-1$ and $a$; steps 11-17) or between
$p_1+1$ and its successor (whose positions are $b-1$ and $b$; steps 19-25). Such a
move is performed only if it is possible to chose $j$ such that, once the block $(p_1\ldots p_j)$ 
is moved, two strip concatenations are possible, at both its ends. If such a transposition is not 
possible, then either $p_1-1$ (steps 27-32) or $p_1+1$ (steps 34-38) allows us to concatenate two strips either by a reversal or by a transposition,
depending on the type of the involved strips.

\medskip

{\em Our implementation.} In our implementation of  Algorithm \ref{algo:1}, the permutation $P$ is stored as a list $L=\{p_1, p_2, \ldots, p_n\}$ 
implemented as a $\log$-list. In addition we need, for each element $p_i$ in $L$, two pointers $b[p_i]$ and $e[p_i]$, 
such that  $b[p_i]$ (respectively $e[p_i])$) is 
not null if and only if $p_i$ is the last (respectively the first) element in its strip (in the order from $\first(L)$ 
to $\last(L)$). In this case, $b[p_i]$ (respectively $e[p_i]$) points to the first (respectively last) element of its strip.
Obviously, these pointers of the abstract data structure $L$ may be directly added to the underlying dynamic tree 
structure $T(L)$.  We note that:

\begin{itemize}
\item[$1)$] the operations $\findelement(L,i)$ and $\findindex(L,x)$ act similarly to $P[i]$ ($=p_i$) and
$P^{-1}[x]$, except that they take $O(\log\, n)$ time, that $\findelement(L,i)$ returns a pointer instead 
of a value  (an operation $\get$ further allows us to obtain this value), and that $\findindex(L,x)$ needs 
a {\em pointer} $t(x)$ to the source of the arc containing $x$
(see Remark \ref{rem:pointerbefore}). The computation of $t(x)$ may be easily performed by noticing that the
nodes are stored only once, when they are created, and the operations on the $\log$-list affect only
the informations at the nodes, and not the nodes themselves. We may therefore store a {\em landmark table} $Lm$ 
with entries $1, 2, \ldots, n$ and such that $Lm[x]$ is always a pointer to one of the endpoints of the
edge with cost $x$, initialized as $t(x)$. Any list-operation modifying the topology of the tree either leaves $t(x)$
unchanged ($\ins$ and $\delete$, except at the boundaries of the inserted/deleted sublist) or puts it on the other endpoint 
of the same edge ($\reverse$, except at the boundaries of the reversed sublist). Moreover, at the
boundaries the number of elements concerned is constant (at most four) and thus updating $Lm[x]$ for these
elements is easy. Then $t(x)$ is either $Lm[x]$ or $\pre(Lm[x])$, and may be computed in $O(\log \, n)$.

\item[$2)$] prefix transpositions and respectively prefix reversals are performed using a $\delete$ and an $\ins$ operation on $L$, 
respectively using a $\reverse$ operation. Both assume the parameters are given as pointers instead of positions,
and need $O(\log\, n)$ time. Updating the strips, {\em i.e.} updating $b[]$ and $e[]$, 
when concatenations hold at the endpoints of the moved or reversed block is also done in $O(\log\, n)$ time. Indeed, at
most two concatenations hold once a block is moved/reversed, and - for each of them - updating needs only to cross the boundary between the  
concatenated strips (with $\pre$ and $\suc$) and to follow the old $b[]$ and $e[]$ values in order to
compute the new values. A prefix transposition or prefix reversal, together with the strips update, thus
takes $O(\log\, n)$ time.
\end{itemize}

With these remarks, it seems that we only succeeded to increase the running time of important operations
instead of reducing it. However, we are able to show that:

\bthm 
Algorithm \ref{algo:1} implemented using $\log$-lists performs in $O(n\, \log\,n)$
time, thus improving the $O(n^2)$ time needed by the algorithm in \cite{dias2015sorting}.
\label{thm:upRT}
\ethm

{\bf Proof.} The initialization of the data structure $H(L)$ is obviously done in $O(n)$ time since the
strips are identified by a simple traversal of the list. The {\em while} loop is executed  $O(n)$ times \cite{dias2015sorting}.
The comparison in line 2 is easily done in $O(\log\, n)$ time 
by testing whether the index of $e[\first(L)]$ is $n+1$ or not.

Steps 4-5 need $O(1)$ time with our implementation, since $i$ is the position of the last element in the
first strip. A pointer $iPtr$ to it is given by $iPtr\leftarrow e[\first(L)]$.

Steps 6-7 need $O(\log\, n)$ time, as $p_1$ is obtained with $\get(\first(L))$ and we have the
two pointers $\first(L)$ and $iPtr$ needed by the transposition (as indicated in $2)$ above).

Steps 9-10 also need $O(\log\, n)$ time, according to $1)$ above.

Steps 11-17 (and similarly 19-25) are also immediate due to 1) and 2). We only have to call
$\findelement(la)$ and $\findelement(a)$ before performing the transposition in  step 14 (and similarly for
the step 17), since in our version of the algorithm we need pointers instead of positions.

The same remarks hold for steps 27-32 (and similarly 34-38). 

The maximum running time of a step is thus of $O(\log\, n)$ implying that the overall
running time is in $O(n\log\, n)$. The theorem is proved. $\Box$
\bigskip

In Figure \ref{fig:Table}, the Unsigned Reversals and Transpositions column records on the
first line the result of Theorem \ref{thm:upRT}. This result may be extended to the two
other columns, and both their lines, since all these four algorithms are based on the
same ideas and have the same bottlenecks as Algorithm \ref{algo:1}: find the last element of the
first strip, compute $P[i]$ ($=p_i$) and $P^{-1}[x]$, and perform transpositions and reversals.
Each of these operations is performed in $O(\log\, n)$ with $\log$-lists, as shown before.
We only have to notice that a signed reversal combines an unsigned reversal and a call to $\changesign$.

The remaining cell in Figure \ref{fig:Table}, which concerns sorting by prefix and suffix variants of transpositions and
of unsigned reversals, is particular. In this case, the algorithm is based on a graph representation of
a permutation, requiring at each step to find a convenient edge, and this is not easier with $\log$-lists
then with classical data structures.
The same reason explains the absence from Figure \ref{fig:Table} of the columns Signed/Unsigned Reversals (only),
for which the implementation with $\log$-lists does not immediately provide an improvement.

\subsection{Replacing permutation trees by $\log$-lists in Sorting by Transpositions}\label{sect:perm}

In \cite{feng2007faster}, Feng and Zhu introduce a new data structure, called permutation
trees, and show that it allows us to improve the running time of the 1.5-approximation algorithm 
for sorting a permutation by transpositions \cite{hartman2006simpler} from $O(n^{\frac{3}{2}}\sqrt(\log\,n))$ time 
to $O(n\log\, n)$ time. Also, the improvement from $O(n^2)$ to $O(n\log\, n)$ time is achieved in \cite{feng2007faster}
for the exact algorithm in \cite{christie1996sorting} for sorting by block interchanges. Recently, 
the 1.375-approximation algorithm \cite{elias20061} for sorting by transpositions has also 
been improved from $O(n^2)$ time to $O(n\log\, n)$ time in \cite{cunha2014faster}, using permutation trees.

Given a permutation (or only a block of it) of size $n$, a permutation tree stores $O(n)$ space
information about it and may be computed in $O(n)$ time. Moreover,  the following operations
are performed in $O(\log\, n)$ time on permutation trees: {\it find the element} at a given position
in the permutation, {\it find the position} of a given element of the permutation, {\it join} the 
permutation trees of two neighboring blocks, {\it split} a permutation tree into two permutation trees corresponding
to a given decomposition of the permutation into two blocks, and {\it query} a given block of 
the permutation represented by the tree looking for the maximum element in the block.

It is easy to see that $\log$-lists also allow to perform all these operations in $O(\log\, n)$ time,
using respectively the operations $\findelement$, $\findindex$, $\ins$, $\delete$ and $\findmax$.
Then we have:

\bthm
$\mlog$-lists successfully replace permutation trees in the $O(n\log\,n)$ time implementations
of the  1.5- and 1.375-approximation algorithms \cite{feng2007faster, cunha2014faster} for sorting by transpositions, 
as well as of the algorithm for sorting by block interchanges \cite{feng2007faster}.
\ethm

\section{Conclusion}\label{sect:conclusion}

The data structure we proposed in this paper has significant interest when compared to existent data structures,
as it combines the advantages of double-linked lists and those of binary search trees, and moreover adds some 
aggregate operations. Like a double-linked list, it allows us to insert/delete/reverse a sublist by cut and link operations.
Like a binary search tree, it allows us to store elements according to their value (or key) order, and search for
the element with a given value, or with the minimum/maximum value, using (basically) a binary search.
In addition, $\log$-lists allow us to keep trace of the rank of each element in its list, and search the element with a given rank.

We proposed here several applications to permutation sorting. They show that the optimal $O(n\log\,n)$ running
time may be achieved for some algorithms whose main challenges are to handle the rank of the elements in the permutation,
to perform a transposition or a reversal, and to merge two parts of the permutation. These operations are simple using
$\log$-lists, as all the difficulties are transferred to a lower abstraction level, handled using link-cut trees.
Link-cut trees already had a lot of the functions needed by $\log$-lists, but still were insufficient without the 
two supplementary operations we added in this paper.

Still, some permutation sorting algorithms use graph representations of the permutation, in which the
search of the best move to perform is difficult mostly due to the graph complexity than to the data structure.
In these cases, $log$-lists still allow to perform the transpositions and reversals on the permutation, but
the running time is not easily improved with $\log$-list. However, as it can be seen even in our applications,
$\log$-lists have a lot of operations, and we think that a more intensive and clever use of aggregate operations 
could allow further improvements.

\bibliographystyle{plain}
\bibliography{CutLink}
\end{document}